\documentclass[12pt,english]{article}
\usepackage{graphicx}

\usepackage{epsf}
\usepackage{amssymb}

\def\dfrac#1#2{{\displaystyle#1\over\displaystyle#2}}

\begin{document}

\title{The Scale Factor in the Universe with Dark Energy}

\author{M.V. Sazhin, O.S. Sazhina, U. Chadayammuri}

\maketitle

\section{Abstract}

Friedmans cosmological equations for the scale factor are analyzed for the Universe containing dark energy. The parameter of the equation of state of the dark energy is treated as an arbitrary constant whose value lies within the interval $w \in [-1.5, -0.5]$, the limits of which are set by current observations. A unified analytic solution is obtained for the scale factor as a function of physical and conformal time. We obtain approximated solutions for scale factor to an accuracy of better then 1\%. This accuracy is better then measurement errors of global density parameters and therefore is suitable for the approximated models of our Universe. An analitic solution is obtained for the scale factor in $\Lambda$CDM cosmological model both in physical and conformal time, for the description of the evolution of the Universe from the epoch of matter domination up to the infinite future.

\section{Introduction}

Since the discovery of the accelerated expansion of our Universe [1],  several theories have been put forward to explain the phenomenon: presence of a cosmological constant, modification of gravity on the largest scales of space-time, presence of new light fields (see, for example, reviews [2, 3, 4, 5, 6, 7]) and a Universe dominated by Chaplygin gas [8]. The substance that produces an effective anti-gravity force has been called dark energy.

Where dark energy is modelled as the manifestation of a new physical field, it can be characterized by the equation of state $p=w\rho $, where the parameter $w$ can differ from $-1$ and, in the general case, is a function of time. In a simple model, where the dark energy is a scalar field with a positive energy (quintessence), the parameter $w$ follows the limitation  $w>-1$, while the cosmological constant corresponds to $w=-1$. However the equation of state can also be strongly negative, $w<-1$; dark energy with such an equation of state is called phantom energy. Current cosmological observations do not exclude the possibility of a time-dependent $w$ where at relatively large values of red shift $z$ the equation of state corresponded to a quintessence with $w>-1$, while in a later epoch it was manifested as a phantom energy with $w<-1$~\cite{Komatsu:2008hk, Sahni:2008xx, Xia:2008ex, Caldwell:2002, lib09}.

The basic function characterizing the global properties of our Universe is the scale factor. An exact equation for this scale factor as a function of physical and conformal time is desirable for the analysis of several problems. This paper is dedicated to the derivation of such a solution and its approximated forms.

The constitution of our Universe is characterized by parameters of density of the individual components: $\Omega_m$ and $\Omega_q$ are densities of non-relativistic matter and dark energy, respectively ~\cite{rub08}.
In the particular case of $w=-1$ we define $\Omega_q = \Omega_{\Lambda}$, where $\Lambda$ means cosmological constant. There are several other components in the Universe, however their contribution to the total density is less than 1\% (at all stages following recombination) and so will be ignored throughout this paper. In other words, a unified analytic solution for the scale factor is obtained for all epochs of the Universe since the moment of matter domination.

We start out by describing the exact solution for the scale factor in the Standard Cosmological Model - $\Lambda$CDM. Such an equation for the scale factor as a function of physical time was arrived to in the textbook ~\cite{kol94}.

We derive the solution for the scale factor as a function of both physical and conformal time. In this we calculate the allowed intervals of conformal time, as well as evaluate points of conformal time crucial to the description of the evolution of the Universe. 

In the second section we study the approximated functions describing the scale factor in the early Universe and the later stage, corresponding to $\Lambda$ -term domination. We also analyze the level of certainty in these limits. 

In the following sections we find an exact implicit equation relating conformal time and the scale factor in a model with an arbitrary constant parameter $-0.5 \ge w \ge -1.5$,in the equation of state of the dark energy, as well as critical moments of time for such a solution. Further, we discuss the approximated solutions.

\section{Exact equations for the scale factor in the Standard Cosmological Model}

Analyzing the background space-time, we shall restrict ourselves to the standard model of the expanding Universe, filled with non-relativistic matter and using the $\Lambda$-term as the source of acceleration of cosmological expansion. The metric of the standard model is that of the a space-flat Universe  ~\cite{rub08}:
\[
ds^2 = dt^2 - a^2(t)d\mathbf{x}^2 .
\]
The scale factor $a(t)$ is determined from the Friedman equation, which can be written in the form:
\begin{equation}
\left( \frac{\dot a(t)}{a(t)} \right)^{2}=H_{0}^{2} \left[ \Omega_m \left(
\frac{a(t_0)}{a(t)} \right)^3 + \Omega_{\Lambda} \right],
\label{fried1}
\end{equation}
where $H_{0}$ is the current value of the Hubble constant, $a(t_0) = a_0 = 1$
in this paper is understood as the current value of the scale factor and a dotted variable indicates a derivative with respect to physical time $t$. Furthermore, we work with a system of units corresponding to $c=1$. In this work we use $\Omega_m=0.27$ and $\Omega_{\Lambda}=0.73$ as the current values of the density parameters~\cite{wmap}. Here we have chosen a set of parameters based on 7-year observations of the WMAP probe and recommended for use. It must be noted that there exist several other models corresponding to different values of the global cosmological parameters. However, for our analysis this difference is not a principal one and a recalculation of the fundamental results is easily accomplished for any choice of these global parameters. 

This standard model of our Universe has been revaluated time and again wince the discovery of its accelerated expansion. 

Our aim is to find an explicit form of the scale factor as a function of physical time $t$ and conformal time $\eta$, defined as:

$$
d\eta = \dfrac{dt}{a(t)}.
$$

We produce these equations and determine the explicit dependence of the scale factor on physical and conformal time, as well as derive several often used equations determining the age of our Universe, moment of transition from decelerated to accelerated expansion, etc.

Friedmans first equation may be written in the form:

\begin{equation}
\frac{1}{a^2(t)}\left( \frac{d a(t)}{dt} \right)^{2}=H_{0}^{2} \left[ \Omega_m
\frac{1}{a^3(t)}  + \Omega_{\Lambda}  \right].
\label{fried2}
\end{equation}

This equation can be integrated to obtain the following relationship between time and the scale factor:

\begin{equation}
H_{0} t = \dfrac{2}{3\sqrt{\Omega_{\Lambda}}} \int\limits_0^{x_s} \dfrac{dx}{\sqrt{1+x^2}}.
\label{scale1}
\end{equation}

Here the upper bound of integration depends on the scale factor as 
\[
x_s = \sqrt{\dfrac{\Omega_{\Lambda}}{\Omega_m}} a^{3/2}.
\]

The current age of the Universe can be determined from this integral by substituting  $a=1$ or $ x_s = \sqrt{\dfrac{\Omega_{\Lambda}}{\Omega_m}} $. The age of the Universe as a function of these global parameters is~\cite{kol94}:

\begin{equation}
H_{0} t_0 = \dfrac{2}{3\sqrt{\Omega_{\Lambda}}} \ln \left(\dfrac{1 + \sqrt{\Omega_{\Lambda}}}{\sqrt{\Omega_m}}\right).
\label{age}
\end{equation}

The solution for the scale factor can be calculated \cite{kol94} from the explicit form of integral (\ref{scale1}):

\begin{equation}
a(t) = \left(\dfrac{\Omega_m}{\Omega_{\Lambda}}\right)^{1/3} \left[\sinh\left( \dfrac{3}{2}\sqrt{\Omega_{\Lambda}} H_{0} t\right) \right]^{2/3}.  
\label{scale2}
\end{equation}

We now analyze the value of the scale factor $a(t)$ at the two limiting cases. The first is the limit of small $H_{0} t << 1$. In this case, the contribution of the $\Lambda$-term to the total density is infinitesimally small and the Universal expands at a decelerating rate, by the law corresponding to matter domination. Then, taking the Taylor expansion of (\ref{scale2}) â we have:

\begin{equation}
a(t) = \left(\dfrac{9}{4}\Omega_m\right)^{1/3} \left( H_{0} t\right)^{2/3},  
\label{scale3}
\end{equation}

which corresponds to the scale factor in the stage of matter domination.

We now calculate the moment of physical time at the beginning of matter domination. This is done by equation of  the parameters of density of matter and relativistic particles (those with an equation of state $\rho = 3p$). The value of physical time at the moment of equal densities of matter and radiation is found from the standard condition ~\cite{rub08}, according to which the scale factor $a_m$ at this moment is equal to the ratio of densities of radiation at the current point of time to the current density of matter:
$$
z_m+1 = \dfrac{a_0}{a_m} = \dfrac{\Omega_m}{\Omega_{\gamma}}
$$

\noindent Hereü $\Omega_{\gamma}$ is the density parameter of relativistic radiation. According to  WMAP~\cite{lar11} the radshift of the epoch is  $z_m = 3196$. Then the moment of physical time $t_m$ corresponding to the epoch of matter domination is:
\begin{equation}
\dfrac{t_m}{t_0} = \left(\dfrac{\Omega_{\Lambda}}{\Omega_m}\right)^{1/2} 
\dfrac{a_m^{3/2}}{\ln \left(\dfrac{1 + \sqrt{\Omega_{\Lambda}}}{\sqrt{\Omega_m}}\right)} .  
\label{age0}
\end{equation}

In the other limit, i.e. for large values of $H_{0} t >> 1$, the contribution of the $\Lambda$-term to the total density becomes dominant, while that of matter significantly diminishes. Now we obtain a different dependence of the scale factor on time:

\begin{equation}
a(t) = \left(\dfrac{\Omega_m}{4\Omega_{\Lambda}}\right)^{1/3} exp\left(\sqrt{\Omega_{\Lambda}} H_{0} t\right),  
\label{scale4}
\end{equation}

\noindent which corresponds to the scale factor of a Universe with a de Sitter expansion law.

We now examine the switch from decelerated to accelerated expansion. To this end we analyze FriedmanÕs second law. In terms of physical time the second equation reads:

\begin{equation}
\frac{1}{a(t)}\left( \frac{d^2 a(t)}{dt^2} \right)^{2}=-{1\over 2}H_{0}^{2} \left[ \Omega_m
\frac{1}{a^3(t)}  - 2\Omega_{\Lambda}  \right].
\label{fried3}
\end{equation}

The moment of time $t=t_{\Lambda}$ when decelerated expansion changes into accelerated expansion is determined by the equation:

\begin{equation}
\frac{d^2 a(t)}{dt^2} = 0,
\label{dec}
\end{equation}

\noindent At this moment of time the value of the scale factor is:

\begin{equation}
a_{\Lambda} =  \left(\dfrac{\Omega_m}{2\Omega_{\Lambda}}\right)^{1/3},
\label{dec1}
\end{equation}

\noindent which corresponds to the red shift:

\begin{equation}
z_{\Lambda} = \dfrac{1}{a_{\Lambda}} -1 = \left(\dfrac{2\Omega_{\Lambda}}{\Omega_m}\right)^{1/3} -1.
\label{dec2}
\end{equation}

The moment of time when the stage of deceleration changes into the stage of acceleration is determined according to:

\begin{equation}
H_{0} t_{\Lambda} = \int\limits_0^{a_{\Lambda}} \dfrac{\sqrt{a}da}{\sqrt{\Omega_m + \Omega_{\Lambda} a^3}}.
\label{age1}
\end{equation}

We calculate the values of physical time at various epochs by for global parameters $\Omega_m = 0.27$, $\Omega_{\Lambda} = 0.73$ $H_0 = 71$ km/s/Mpc. The moment of matter domination corresponds to value of  red shift $z=3196$ and constitutes 98357 years from the beginning of expansion. The moment of change from deceleration to acceleration is $t_{\Lambda}=$7.1 billion years, and the red shift at this point has a value of $z_{\Lambda}=0.82$. Finally, the age of the Universe constitutes 13.75 billion years, with the current value of red shift equal to  $z=0$. The Universe infinitely in the future corresponds to a physical time $t=\infty$.

The dependence of the scale factor on physical time is not sufficient for several formulae. To be used in these formulae, it must be expressed as a function of conformal time. This is most easily obtained by rewriting FriedmanÕs first equation in conformal time:

\begin{equation}
\frac{1}{a^4(\eta)}\left( \frac{d a(\eta)}{d\eta} \right)^{2}=H_{0}^{2} \left[ \Omega_m
\frac{1}{a^3(\eta)}  + \Omega_{\Lambda}  \right].
\label{fried4}
\end{equation}

\noindent The solution to this equation in integral form is:
\begin{equation}
H_{0} \eta = \int\limits_0^{a} \dfrac{dx}{\sqrt{x}\sqrt{\Omega_m + \Omega_{\Lambda} x^3}}.
\label{age2}
\end{equation}

This integral can be expressed in explicit form as an elliptical integral $F(\varphi, k)$ of the first kind.  For this, the integral is rearranged as follows:

\begin{equation}
\Omega^{1/6}_{\Lambda} \Omega^{1/3}_m H_{0} \eta = \int\limits_0^{u} \dfrac{dx}{\sqrt{x}\sqrt{1 +  x^3}}.
\label{scale5}
\end{equation}

\noindent The upper limit of integration here is $u=\left(\dfrac{\Omega_{\Lambda}}{\Omega_m}\right)^{1/3}a$. The integral (\ref{scale5}) is solved ~\cite{gr63} as:

\begin{equation}
3^{1/4} \Omega^{1/6}_{\Lambda} \Omega^{1/3}_m H_{0} \eta = F\left(\arccos\dfrac{1+(1-\sqrt{3})u}{1+(1+\sqrt{3})u}, \dfrac{\sqrt{2+\sqrt{3}}}{2}\right).
\label{scale6}
\end{equation}

Elliptic integrals can be inverted using Jacobi elliptic functions to obtain an explicit formulation of the scale factor as a function of conformal time. Thus we use the elliptic cosine ~\cite{htf} to describe the scale factor function:

\begin{equation}
a(\eta) = \left( \dfrac{\Omega_m}{\Omega_{\Lambda}}\right)^{1/3} \dfrac{1 - cn(y, k)}{\sqrt{3}(1 + cn(y, k)) - (1 - cn(y, k))}. 
\label{scale7}
\end{equation}

\noindent where the argument of the cosine is $y=3^{1/4} \Omega^{1/6}_{\Lambda} \Omega^{1/3}_m H_{0} \eta$ and its modulus is $k=\dfrac{\sqrt{2+\sqrt{3}}}{2} \approx 0.97$.

Now we need to determine the moments of conformal time corresponding to the following epochs: the begin on expansion $\eta_s$, the beginning of matter domination $\eta_m$, the epoch of change from decelerated to accelerated expansion $\eta_{\Lambda}$, the current epoch $\eta_0$ and the infinite future $\eta_{\infty}$. Since conformal time does not have a direct physical meaning, the way physical time does, we will be calculating not  $\eta$ itself, but the two quantities  $y$ and $H_0\eta$. It must be noted that the quantity $y$ can be evaluated exactly, while $H_0\eta$ is expressed through $y$ and the two observationally determined constants (density parameters) $\Omega_{\Lambda}, \quad \Omega_m$, that have a precision of a few percent. The precision of $H_0\eta$ is thus constrained by that of the density parameters, and we state the numerical values of this quantity up to only two decimal points. It follows that our model is inadequate for describing the early Universe, i.e. this model gives an inexact law of evolution of the scale factor in the interval from the beginning of expansion to the moment of matter domination. However, after this moment our solutions get more accurate as the point of time in question moves closer to the current moment. Thus, while we so analyze evolution starting at the beginning of expansion, the solution is considered qualitative until the moment of matter domination.

At the beginning of expansion the scale factor has a value of zero, which is thus the lower bound for allowed values of conformal time. In other words, at the beginning of expansion $y=0$, $H_0\eta_s = 0$. 

The value of conformal time at the moment of equal densities of matter and radiation is found from the standard condition, as in the case of physical time. This gives us the value for the variable $ y_m = 0.05$ and the value of conformal time at this moment:
\begin{equation}
H_0\eta_m = 6.2 \cdot 10^{-2}.
\label{timem}
\end{equation}

This moment of time we shall henceforth call the beginning of the stage of matter domination. 

We are now left to calculate conformal times corresponding to the switch from decelerating to accelerating expansion, the current epoch and the infinite future. The moment of time corresponding to (\ref{dec1}) is:
$$
cn(y_{\Lambda}, k) = \dfrac{1+2^{1/3} -3^{1/2}}{1+2^{1/3} +3^{1/2}}.
$$

\noindent The solution to this equation is $y_{\Lambda} = 2.27$, which for the accepted values of the global parameters is equivalent to:\begin{equation}
H_0 \eta_{\Lambda} = 2.82.
\label{timed}
\end{equation}

The current epoch is obtained by setting the scale factor to one. This leads to the solution:

$$
cn(y_0, k) = \dfrac{1 - \left(\dfrac{\Omega_{\Lambda}}{\Omega_m}\right)^{1/3}(\sqrt{3} - 1)}{1 + \left(\dfrac{\Omega_{\Lambda}}{\Omega_m}\right)^{1/3}(\sqrt{3} + 1)}.
$$

\noindent For arbitrary values of the density parameters this solution can be rewritten as $y_0 = 2.78$. Alternatively, for the selected values of these global parameters we have:

\begin{equation}
H_0 \eta_0 = 3.45.
\label{time0}
\end{equation}

The moment of time in the infinite future corresponds to the scale factor going to infinity (in the standard cosmological model the Universe expands infinitely). Correspondingly the first zero of the denominator in equation (\ref{scale7}) determines the moment of conformal time representing the infinite future. This solution is:

$$
cn(y_{\infty}, k) = \dfrac{1 -3^{1/2}}{1 +3^{1/2}}.
$$

\noindent Again, the solution to this equation for arbitrary values of the density parameters is $y_{\infty} = 3.69$ and for the selected values gives:
\begin{equation}
H_0 \eta_{\infty} = 4.57.
\label{timei}
\end{equation}

Thus the complete interval for which $y$ can be defined is:
$$y \in [0, 3.69] ,$$ 
which leads to the interval of allowed values of $H_0 \eta$ being:   
$$H_0 \eta \in [0, 4.57]. $$ 

\noindent The allowed interval of values of conformal time is thus established.

\section{Approximate solutions, error analysis}

In discussing approximated solutions to the scale factor, it must be immediately noted that these approximations are justified only for $z<<10^5$, i.e. at or after the stage of matter domination. At earlier stages the Universe is dominated by radiation, a scalar field of inflaton and possibly yet-unknown forms of matter. Therefore, all our calculations for the real Universe as justified in the interval $H_0 \eta \in [0.0122, 4.57]$. Nevertheless, within this interval of conformal time we may use either the exact or the approximated equations for the scale factor, as is discussed in this section. 

Given an exact solution to FriedmanÕs equations for a scalar factor (in terms of conformal time) we find an approximation to the scale factor in the limit of small values of $a$. Small values of $a$ correspond to $H_{0} \eta << 1$.  The elliptic cosine upon Taylor expansion has the form(~\cite{htf} pg.46, f.11):
\begin{equation}
cn(y, k) = 1 - \dfrac{y^2}{2!} + (1 + 4k^2)\dfrac{y^4}{4!} - (1 + 44 k^2 + 16 k^4) \dfrac{y^6}{6!} + ...
\label{ecos}
\end{equation}

\noindent Leaving terms up to the quadratic in y we obtain the approximate expression for the scale factor:

\begin{equation}
\tilde a(\eta) = \dfrac{\Omega_m}{4} \left( H_{0} \eta \right)^2,
\label{scale8}
\end{equation}

\noindent which corresponds to the expression for the scale factor in the stage of matter domination in terms of conformal time. 

Now we analyze the asymptote of $a(\eta)$ as it approaches $\eta_{\infty}$. For this we define $y=y_{\infty} + \psi$ and Taylor expand the elliptic cosine about  $y_{\infty}$. We first use the formula from addition theorems of the elliptic function: 
\begin{equation}
cn(y_{\infty} + \psi, k) = \dfrac{cn(y_{\infty}, k) cn(\psi, k) - sn(y_{\infty}, k) dn(y_{\infty}, k) sn(\psi, k) dn(\psi, k)}
{1- k^2 sn^2(y_{\infty}, k) sn^2(\psi, k)}.
\label{ellipt}
\end{equation}
 
\noindent Here, $sn(y, k)$ is the elliptic sine, while $dn(y, k)$ is the third elliptic function defined by Jacobi ~\cite{htf}. 

Using the properties of the Jacobi elliptic functions we determine the values of the three functions at the point $y_{\infty}$, In the calculations below we shall not explicitly write out the modulus $k$, and assume its value to be $k= \dfrac{\sqrt{2 + \sqrt{3}}}{2}$ everywhere. 
$$
cn(y_{\infty}) = \dfrac{1 - \sqrt{3}}{1 + \sqrt{3}},
$$

$$
sn(y_{\infty}) = \sqrt{\dfrac{2 \sqrt{3}}{2 + \sqrt{3}}},
$$

$$
dn(y_{\infty}) = \sqrt{\dfrac{2 - \sqrt{3}}{2}}.
$$

Leaving the highest order term of $\psi$ in the Taylor expansion of both the numerator and denominator in (18), we obtain the following expression for the scale factor:
$$
a(\psi) = -\left(\dfrac{\Omega_m}{\Omega_{\Lambda}}\right)^{1/3} \dfrac{3^{1/4}}{\psi}
$$

\noindent Plugging in the expression for $\psi$ in terms of $\eta$ we finally obtain:

\begin{equation}
a(\eta) = \dfrac{1}{\sqrt{\Omega_{\Lambda}}H_0\left(\eta_{\infty} - \eta\right)},
\label{scale9}
\end{equation}

The time variable can be redefined by introducing the new variable $\hat\eta = \eta - \eta_{\infty} $. Then, the quantity $\hat\eta$ appears negative, and the scale factor as a function of $\hat\eta$ is:
$$
a(\hat\eta) = -\dfrac{1}{\sqrt{\Omega_{\Lambda}}H_0\hat\eta},
$$
\noindent which corresponds to the known solution for the scale factor (as a function of conformal time) in a de Sitter metric~\cite{rub08}. 

We now analyze errors that occur when evaluating the scale factor through formulae (\ref{scale8}) and (\ref{scale9}) and matching together  these approximated solutions, for example, at $a \approx 1$. We start with the approximation (\ref{scale8}). Evaluation of errors of the approximate solutions is done by plugging in (\ref{scale8}) at the moments of time evaluated in (\ref{timed}) and (\ref{time0}). The error evaluation at the remaining points of time is trivial. At the point (\ref{timem}) the approximate and exact values of the scale factor are equal with an accuracy of greater than 1\% while at the point (\ref{timei}) the ratio of the approximated scale factor to the exact value equals zero.

We first evaluate the accuracy of extrapolating the lower-limit approximation (\ref{scale8}) to higher values of conformal time:
$$
\dfrac{\tilde{a_{\Lambda}}}{a_{\Lambda}} = 2^{1/3} \dfrac{(1.73)^2}{4} \approx 0.94,
$$

$$
\dfrac{\tilde{a_0}}{a_0} = \dfrac{(2.12)^2}{4} \left(\dfrac{\Omega_m}{\Omega_{\Lambda}}\right)^{1/3} \approx 0.81.
$$

\begin{figure*}
\begin{center}
\includegraphics[width=15.0cm]{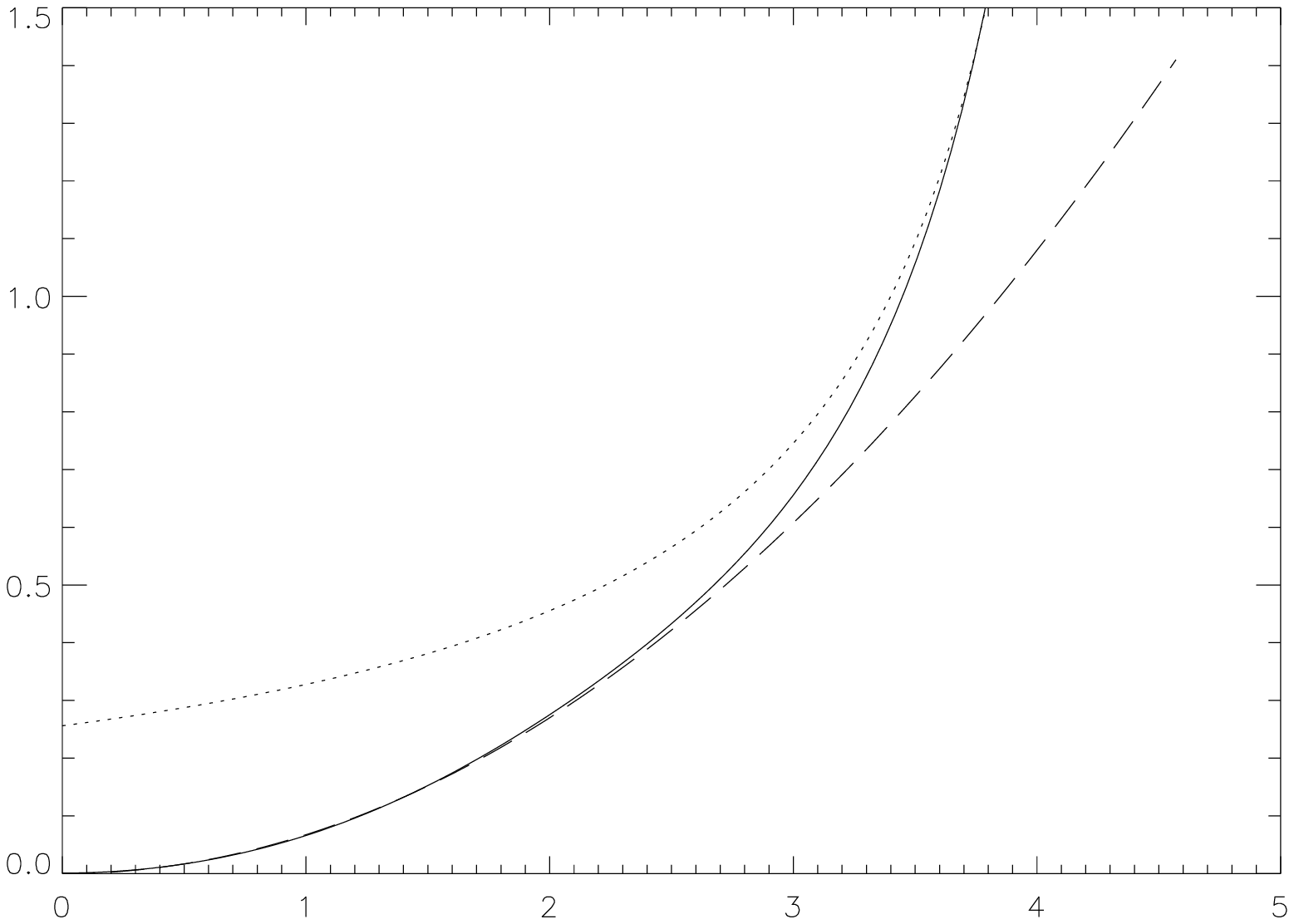}
\end{center}
\caption{The diagram shows the curves describing the evolution of the scale factor. Values of the scale factor are placed along the vertical axis and $H_0\eta$ grows along the horizontal axis. The exact value of the scale factor is represented by the solid line, the approximated formula (\ref{scale9}) is graphed by the dotted line, and the dashed line shows the scale factor as evaluated from formula (\ref{scale8}). See text for details.}
\label{fig1}
\end{figure*}

It is now seen that the approximated curve described by formula (\ref{scale8})lies under the curve described by the exact formula. The deviation between the two constitutes 6 \% at the point  $H_0 \eta_{\Lambda}$ (See Fig.\ref{fig1})but increases to 19\% at $H_0 \eta_0$.

Calculating the accuracy of interpolation of the upper-limit approximation (\ref{scale9}) into lower conformal times, we have:
$$
\dfrac{\hat{a_{\Lambda}}}{a_{\Lambda}} = 2^{1/3} /1.07 \approx 1.18
$$

$$
\dfrac{\hat{a_0}}{a_0} = \dfrac{1}{0.68} \left(\dfrac{\Omega_m}{\Omega_{\Lambda}}\right)^{1/3} \approx 1.06
$$

Here the approximated curve described by formula (\ref{scale9}), lies above the curve described by the exact formula. At $H_0 \eta_0$  the deviation is 6\% while it increases to reach 18\% at $H_0 \eta_{\Lambda}$.

Since the error upon extrapolation from small values of the scale factor that that upon interpolation from bigger values have different signs upon comparing with the exact value, the errors add up. Therefore the error upon lacing together the scale factor using these two approximated formulae constitutes almost 25\%. 

An even greater error appears upon comparing the first and second derivatives of the scale factor. The first derivative of the logarithm of the scale factor from the side of small vales is two times bigger than the analogous derivative from the side of greater values of the scale factor. A comparison of the second derivatives gives a ratio of values:
$$
{\left(\dfrac{\tilde{a^{\prime\prime}}}{a} \right)\over \left(\dfrac{\hat{a^{\prime\prime}}}{a}\right) } \approx 0.54,
$$
\noindent here prime denotes a derivative with respect to conformal time.
Thus the approximated equations for the scale factor should only be allowed in in calculations that require accuracy no greater than 25\%.

The approximate value of the scale factor can be analyzed close to $a=1$. However, approximating the value of $a$ close to unity by power functions is inefficient, since a large number of terms of the Taylor expansion would be required to obtain an accuracy comparable to 1\%.

Therefore, we shall use another approximate solution for elliptic functions of the form:
$$
cn(y)=\dfrac{2\pi}{k\textbf{K}} \sum\limits_{n=1}^{\infty}\dfrac{q^{n-1/2}}{1+q^{2n-1}} \cos(2n-1)\frac{\pi y}{2\textbf{K}},
$$

\noindent Where $\textbf{K}$ is the complete elliptic integral of the first kind along modulus  $k$, the auxiliary function $q=exp( -\dfrac{\pi \textbf{K}^{\prime}}{\textbf{K}} )$,and the primed value is the complete elliptic integral of the first kind from the remaining value of the modulus.

The first three terms of this expansion already allow for errors less than 0.5\% across the entire interval of $\eta$. Therefore, if the problem lies in finding an approximate solution for the scale factor with accuracy of the order of 1\%, it is sufficient to use the sum of these first three terms of expansion. The value of the complete elliptic integrals is $\textbf{K} = 2.768$, and $\textbf{K}^{\prime}=1.598$, while $q=0.163$.

One can represent the scale factor as:
\begin{equation}
a(\eta) = a_1 \dfrac{a_n(\eta)}{a_d(\eta)},
\label{scale_cur}
\end{equation}

\noindent here

$$
a_1 = \dfrac{\sqrt{3}-1}{2}\left(\dfrac{\Omega_m}{\Omega_{\Lambda}}\right)^{1/3}
$$

$$
a_n(\eta) = 1 - \dfrac{2\pi\sqrt{q}}{k \textbf{K}} \left( \dfrac{1}{1+q}\cos \dfrac{\pi y}{2\textbf{K}} +  \dfrac{q}{1+q^3}\cos \dfrac{3\pi y}{2\textbf{K}} + \dfrac{q^2}{1+q^5}\cos \dfrac{5 \pi y}{2\textbf{K}}\right)
$$

$$
a_d(\eta) = 2 - \sqrt{3} + \dfrac{2\pi\sqrt{q}}{k \textbf{K}} \left( \dfrac{1}{1+q}\cos \dfrac{\pi y}{2\textbf{K}} +  \dfrac{q}{1+q^3}\cos \dfrac{3\pi y}{2\textbf{K}} + \dfrac{q^2}{1+q^5}\cos \dfrac{5 \pi y}{2\textbf{K}}\right)
$$

let introduce variable $\zeta=\sqrt{\Omega_{\Lambda}}H_0 \eta$ and we get the approximate form of scale factor:

\begin{equation}
a(\eta) = 0.263 \dfrac{1 - 0.8159\cos 0.5361 \zeta -  0.1541\cos 1.6083 \zeta - 0.0252\cos2.6805 \zeta}{0.2680 + 0.8159\cos 0.5361 \zeta -  0.1541\cos 1.6083 \zeta + 0.0252\cos2.6805 \zeta},
\label{scale_cur1}
\end{equation}

This equation approximates exact equation (\ref{scale7}) with accuracy better then 1\% over the whole interval of allowed values of $\eta$.

\section{General Case of Parameter $w$}

Analysis of observational data indicates that parameter $w$ is close to minus unity ($w=-1$) \cite{lar11}, \cite{con11}, \cite{san10}. An interval of  acceptable value of $w$ changes significantly depending of the analysis method, observational errors, and bias. Therefore the value of $w$ can be different from -1. Analysis of cosmological models with $w \neq -1$ is requeired. 

We now look at the dependence of the scale factor on time for an arbitrary value of $w \in [-1.5, -0.5]$. This interval of values is slightly larger than that set by observational limits, however, we use it for convenience.

The equation for conformal time $\eta$ as a function of the scale factor a can be obtained from the inverse function. The solution $\eta(a)$ is: 
\begin{equation}
H_{0} \eta = \int\limits_0^{a} \dfrac{dx}{\sqrt{x}\sqrt{\Omega_m + \Omega_q x^{3|w|}}}
\label{age3}
\end{equation}

\subsection{Analitical solution for the scale factor}

Finally, we write out the analytic function describing the conformal time as a function of the scale factor:
\begin{equation}
\frac{1}{2} \sqrt{\Omega_m} H_{0} \eta = \sqrt{a} F\left(\dfrac{1}{6|w|}, \frac{1}{2};\dfrac{6|w|+1}{6|w|}; -\dfrac{\Omega_q}{\Omega_m}a^{3|w|}\right)
\label{scale11}
\end{equation}

\noindent Here $F(\alpha, \beta;\gamma; z)$ is the Gauss hypergeometric function. This equation must be inverted to obtain the equation for the scale factor as a function of conformal time. However, the authors are unaware of special functions that could solve this problem. Therefore, where such a solution is required, it is determined by numerical methods. Fig.\ref{fig4} shows the form of the function $a(\eta)$ for values of the parameter $w=-0.5, -1.0, -1.5$.

Now we find the asymptotic behaviour of the at small and large values of conformal time. It must be noted that the first of these asymptotes must correspond to a matter dominated Universe, and the latter to a Universe dominated by dark energy.

The Gauss hypergeometric function can be Taylor expanded on the argument $-\dfrac{\Omega_q}{\Omega_m}a^{3|w|}$; in the case where the absolute value of this argument is less than unity, this expansion converges. We note that:
$$
F\left(\dfrac{1}{6|w|}, \frac{1}{2};\dfrac{6|w|+1}{6|w|}; 0\right) = 1.
$$

\noindent This means that for small values of the scale factor Ð and thus conformal time Ð we once again obtain formula (\ref{scale8}):
$$
a(\eta) = \dfrac{\Omega_m}{4} \left( H_{0} \eta \right)^2,
$$

\noindent which describes the evolution of the scale factor in the stage of matter domination.

In the second case, where $a \gg 1$, functional relations of hypergeometric functions may be used to obtain a series expansion of the function $F\left(\alpha, \beta; \gamma; z\right)$ in terms of hypergeometric functions of the variable $1/z$~\cite{htf}:

\[
F(\alpha, \beta; \gamma; z) = \dfrac{\Gamma(\gamma)\Gamma(\beta - \alpha)}{\Gamma(\beta)\Gamma(\gamma - \alpha)}\left(-z\right)^{\alpha} F(\alpha, \alpha - \gamma + 1; \alpha - \beta  + 1; \dfrac{1}{z}) + 
\]
\[
+ \dfrac{\Gamma(\gamma)\Gamma(\alpha-\beta)}{\Gamma(\alpha)\Gamma(\gamma - \beta)}\left(-z\right)^{\beta} F(\beta, \beta - \gamma + 1; \beta - \alpha + 1; \dfrac{1}{z}).
\]
 
Using this expansion and the equivalence of hypergeometric functions to unity for an argument approaching zero: $1/z =  -\dfrac{\Omega_m}{\Omega_q} a^{-3|w|} \rightarrow 0$, we obtain the following expression in the limiting case of a large scale factor:

\begin{equation}
H_{0} \eta = H_{0} \eta_{\infty} - \frac{2}{3|w|-1} \frac{1}{\sqrt{\Omega_q}} a^{\frac{1 - 3|w|}{2}},
\label{age5}
\end{equation}

\noindent Here $H_{0} \eta_{\infty}$ is determined by formula (\ref{age4}). Meanwhile, the scale factor is now determined by the expression:

\begin{equation}
a(\eta) = \dfrac{1}{
\left(\frac{3|w|-1}{2}\sqrt{\Omega_q}H_0 (\eta_{\infty} - \eta)\right)^{\frac{2}{3|w|-1}} 
}
\label{scale12}
\end{equation}

It may also be noted than for $w=-1$, the equation (\ref{scale11}) reduces to:
$$
a(\eta) = \dfrac{1}{
\sqrt{\Omega_q}H_0 (\eta_{\infty} - \eta)
}
$$

Although the set of exact results relating to the evolution of the scale factor got an arbitrary value of parameter $w$ appears limited in completeness, certain interesting formulations may be suggested for the implicit expression of the scale factor through conformal time. 

The hypergeometric function of the given values of parameters may be expressed in terms of an incomplete beta function. This function is used in mathematical statistics. Some minor standard reformulations also allow us to express the hypergeometric function in terms of Legendre functions.

We produce the equation for the scale factor in a form convenient for the analysis of its evolution in the limit $a >> 1$:
\begin{equation}
\dfrac{a^{\frac{3|w|-1}{2}}}{F\left(\frac{1}{2}, \frac{1}{2}-\frac{1}{6|w|}; \frac{3}{2}-\frac{1}{6|w|}; \frac{\Omega_m}{\Omega_q a^{3|w|}}\right)}
=\frac{2}{3|w|-1}
\dfrac{1}{\sqrt{\Omega_q} H_0  \left(\eta_{\infty} - \eta\right)^{\frac{2}{3|w|-1}} }
\label{scale13}
\end{equation}

Let us consider the conformal time intervals. It is easily shown that this integral (\ref{age3}) has a finite value for infinite limits under the condition $3|w| > 1$. Otherwise, the value of the integral (\ref{age3}) diverges at infinity. In other words, in the interval of values of $|w|$ of interest to us the integral (\ref{age3}) is finite and thus so is the finite interval of values of  $\eta$.

The value of conformal time for which the scale factor goes to infinity is:

\begin{equation}
H_{0} \eta_{\infty} = \dfrac{1}{3|w|\sqrt{\pi \Omega_m}} \left(\dfrac{\Omega_m}{\Omega_q}\right)^{\frac{1}{6|w|}} \Gamma\left(\dfrac{1}{6|w|}\right)\Gamma\left(\dfrac{3|w|-1}{6|w|}\right).
\label{age4}
\end{equation}

The complete interval of conformal time depends on the parameter $w$. For an upper limit of the parameter $w = -0.5$ the interval of change of conformal time is $H_0\eta_{\infty} \in [0, 7.75]$. For the lower bound $w = -1.5$, this interval is $H_0\eta_{\infty} \in [0, 4.20 ]$.

Let us estimate the conformal time at the moment $H_{0} \eta_q$ of zero decelaration. 

This moment is:
\begin{equation}
\frac{1}{2}\sqrt{\Omega_m}H_{0} \eta_q = \dfrac{1}{(3|w|-1)^{1/6|w|}} \left(\dfrac{\Omega_m}{\Omega_q}\right)^{\frac{1}{6|w|}} F\left(\dfrac{1}{6|w|}, \dfrac{1}{2}; \dfrac{6|w|+1}{6|w|};  -\dfrac{1}{3|w|-1}\right).
\label{age5}
\end{equation}

The graph of $H_{0} \eta_q$ as function of argument $w$ is shown in Fig.\ref{fig5}.

\subsection{Measuring the parameter $|w|$.}

The red shift at the moment of time when deceleration turned to acceleration, in this general case, corresponds to:
\begin{equation}
z_q = \dfrac{1}{a_q} -1 = \left((3|w|-1)\dfrac{\Omega_q}{\Omega_m}\right)^{1/3|w|} -1,
\label{dec3}
\end{equation}

Measurement of the parameter of the equation of state poses a tough but very important problem. We focus on one possibility of measuring $|w|$ from measurements of red shift $z_q$. Measuring the value of red shift upon which decelerated expansion shifts to an accelerated one, it is possible to measure the parameter $w$. It is interesting to note that the dependence $z_q \div |w|$ is not monotonous. We also note that the maximum of the curve in Fig. \ref{fig3} corresponds to the solution to the equation:
$$
exp \biggl( \dfrac{3|w|}{3|w|-1} \biggr) = (3|w|-1)\dfrac{\Omega_q}{\Omega_m},
$$
\noindent and consequently there exists a unique relationship between the parameter $w$ and the maximal value of $z_q$.

To estimate the errors of $w$ parameter we have to evaluate the derivative:
\begin{equation}
\dfrac{dz_q}{d|w|} = -\dfrac{1+z_q}{3w^2} \left(\ln(3|w|-1) - \frac{3|w|}{3|w|-1} + \ln\frac{\Omega_m}{\Omega_q}\right).
\label{derv}
\end{equation}

It is possible to try to approximate the solution for a scale factor with arbitrary $w$ to that of the standard model of the Universe. To this end we determine the function with an accuracy of:
\begin{equation}
\Delta = \dfrac{a(\eta, w) - a(\eta, w=-1)}{a(\eta, w=-1)},
\label{exact}
\end{equation}

\noindent The graph of this function if shown in Fig.\ref{fig6}. It is observed that the approximation varies from the exact solution by over 10\%. This means that the solution to the scale factor in the standard model of the Universe is not suitable to describe the evolution of a Universe filled with dark matter with an arbitrary parameter $w$.

\section{Conclusion}

In this paper we analyzed exact solutions for the scale factor to model a Universe filled with matter and dark energy. 

We first comprehensively analyzed exact solutions for the scale factor in the standard model of the Universe filled with dust-like matter and containing a $\Lambda$-term. Exact solutions are derived and presented in terms of physical and conformal time. The latter exact solution, in conformal time, is expressed in the form of Jacobi elliptic functions. We consider it to be particularly important, since the analysis of fluctuations of the gravitational field is conducted in terms of conformal time. We also studied approximated solutions in the standard model of the Universe. It was shown that approximated solutions in the form of power laws were not satisfactory. Their inaccuracy significantly exceeds the error in determining fundamental global parameters of our Universe. Thus, it is impossible to use these approximated solutions for a majority of problems. We found an expression for the exact solution as a sum of three trigonometric functions, which comprise a solution for the scale factor with an accuracy better than 1\% in the entire interval of allowed values for conformal time. Such an approximation may be used for all problems, since the accuracy of the approximation is higher than the rms of global cosmological parameters.

We also found exact solutions for conformal time as a function of the scale factor for arbitrary values of the parameter $w$. A satisfactory approximation of this solution could not be found in terms of reasonably simple functions, although certain useful approximations are discussed in the paper.

The paper also discusses the entire allowed interval of values of conformal time for a Universe filled with dark energy with varying values of the parameter $w$, as well as certain other critical points of time: the moment of the beginning of matter domination, the moment of change from decelerated to accelerated expansion and the moment which corresponds to the infinite value of the scale factor.
 
\section{Acknowledgements}
The authors are indebted to A.O. Marakulin for the accistance in some calculations. The research was financially supported by the grant RFFI 10-02-00961a, grant MKÊ-473.2010.2. of the President of the Russian Federation, and the International Scholars Program of Brown University, Providence, Rhode Island, USA. The work was carried out as part of the project No. 14.740.11.0085 of the Ministry of Education.

\clearpage

\newpage

\begin{figure*}
\begin{center}
\includegraphics[width=15.0cm]{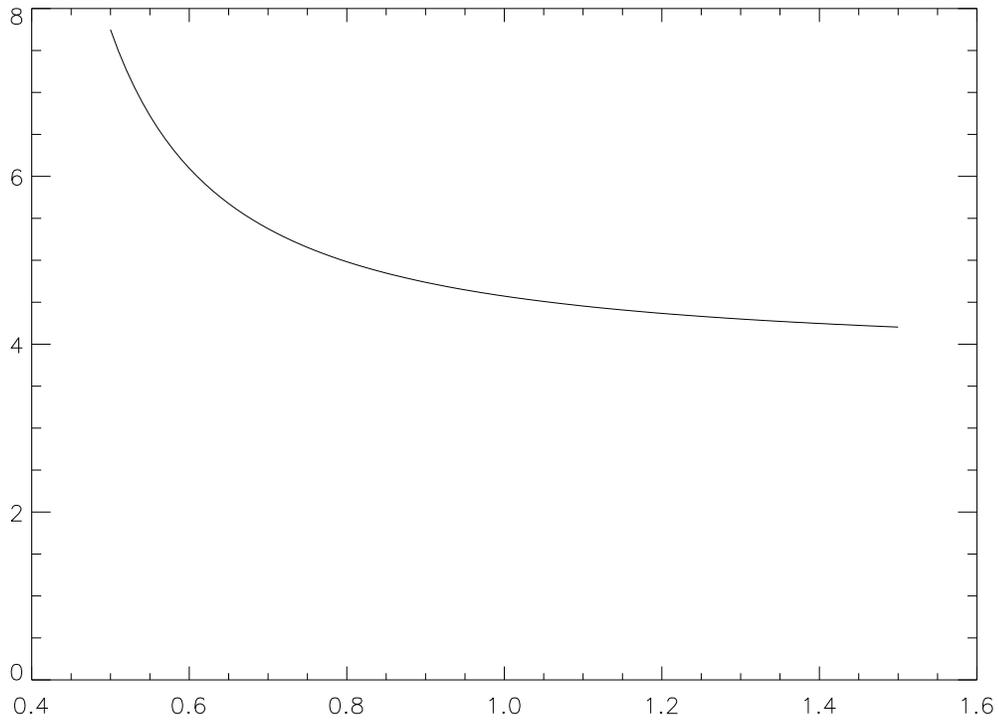}
\end{center}
\caption{The diagram shows the dependence of  $H_0\eta_{\infty}$ on the parameter $w$. it is observed that when the parameter is in the interval $-0.8 \le w \le -0.5$ the value of $H_0\eta_{\infty}$ varies strongly with the parameter. After -0.8 up to the end of the allowed interval of values for the parameter, $H_0\eta_{\infty}$ depends on it weakly, with its value staying between 4.98 and 4.2.}
\label{fig2}
\end{figure*}

\begin{figure*}
\begin{center}
\includegraphics[width=15.0cm]{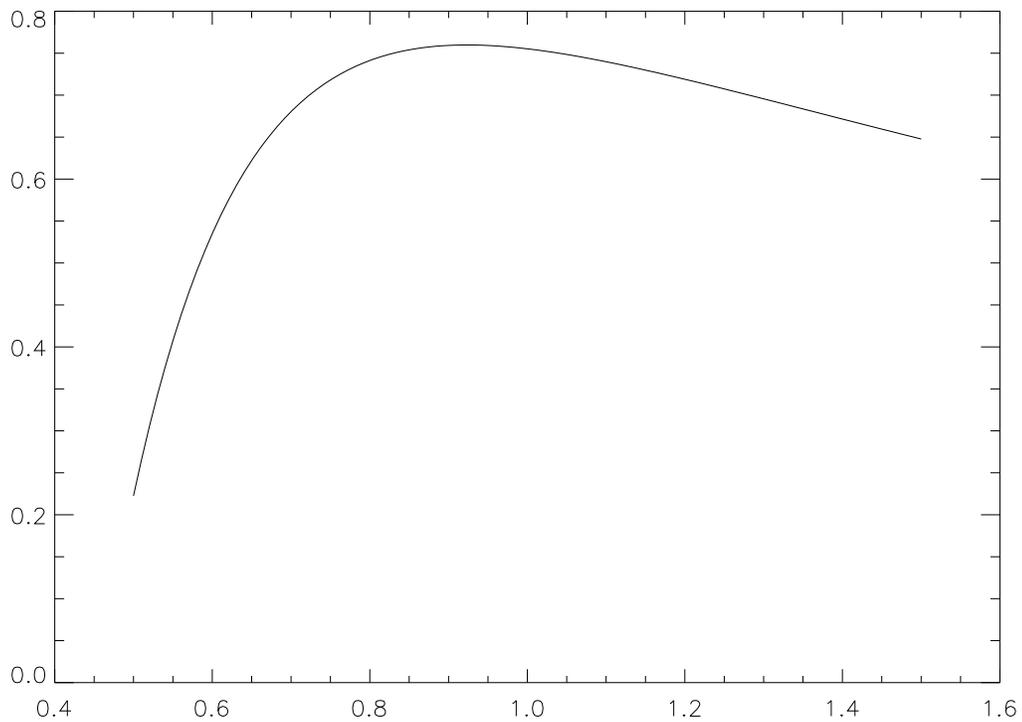}
\end{center}
\caption{This diagram shows the value of red shift $z_q$ upon the change from decelerated to accelerated expansion. The argument of this function is the parameter of equation of state $|w|$. It is interesting to note the non-monotonous dependence of $z_q$ on $w$. The maximum value of  $z_q$ corresponds approximately to $w=-0.9$.
.}
\label{fig3}
\end{figure*} 

\begin{figure*}
\begin{center}
\includegraphics[width=15.0cm]{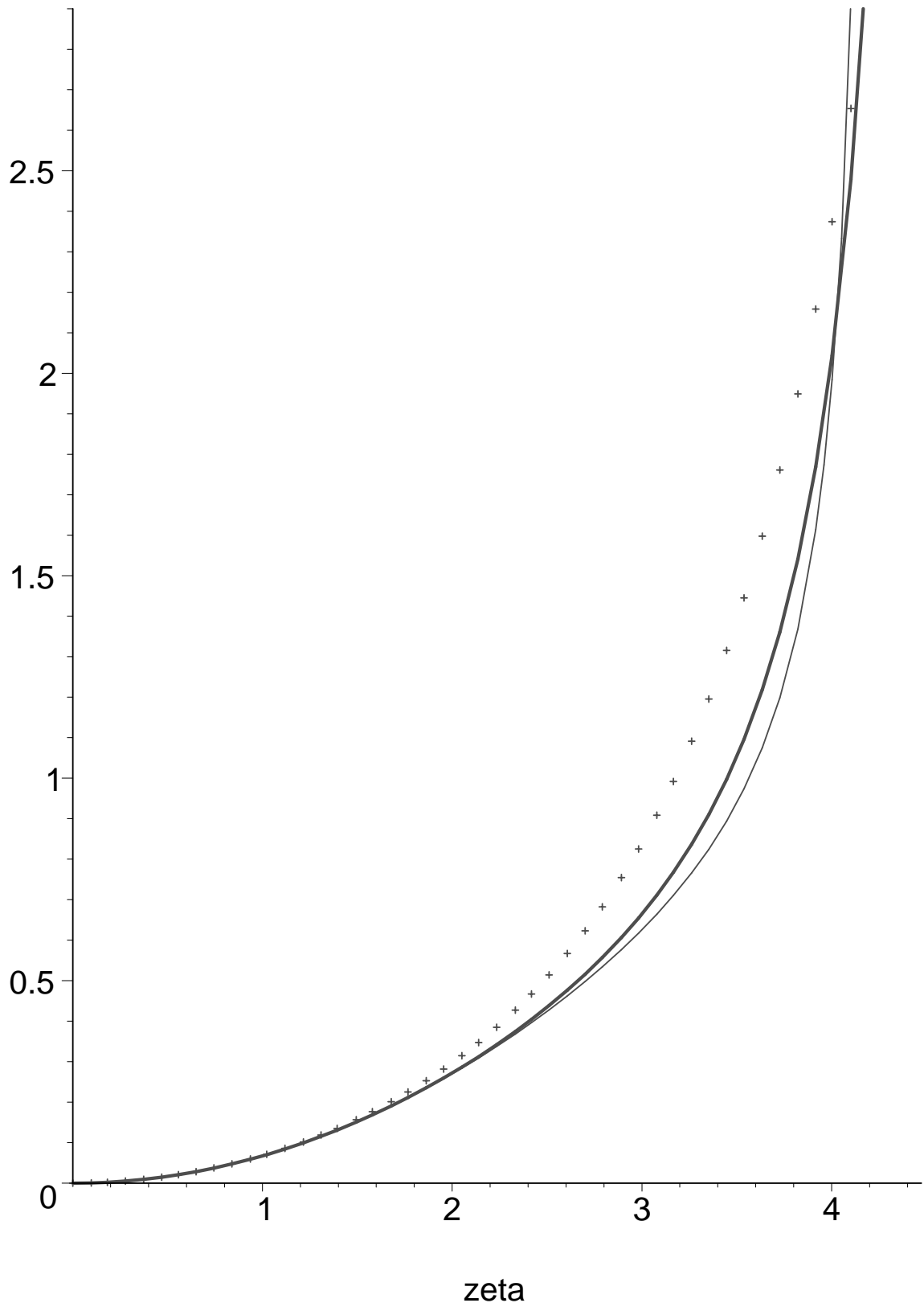}
\end{center}
\caption{Here scale functors as a function of $H_0\eta$ are plotted for three values ofdark energy parameter $w$. Bold line corresponds $w=-1$, asterics correspond to -- $w=-0.5$, and light  line corresponds to -- $w=-1.5$.}
\label{fig4}
\end{figure*}

\begin{figure*}
\begin{center}
\includegraphics[width=15.0cm]{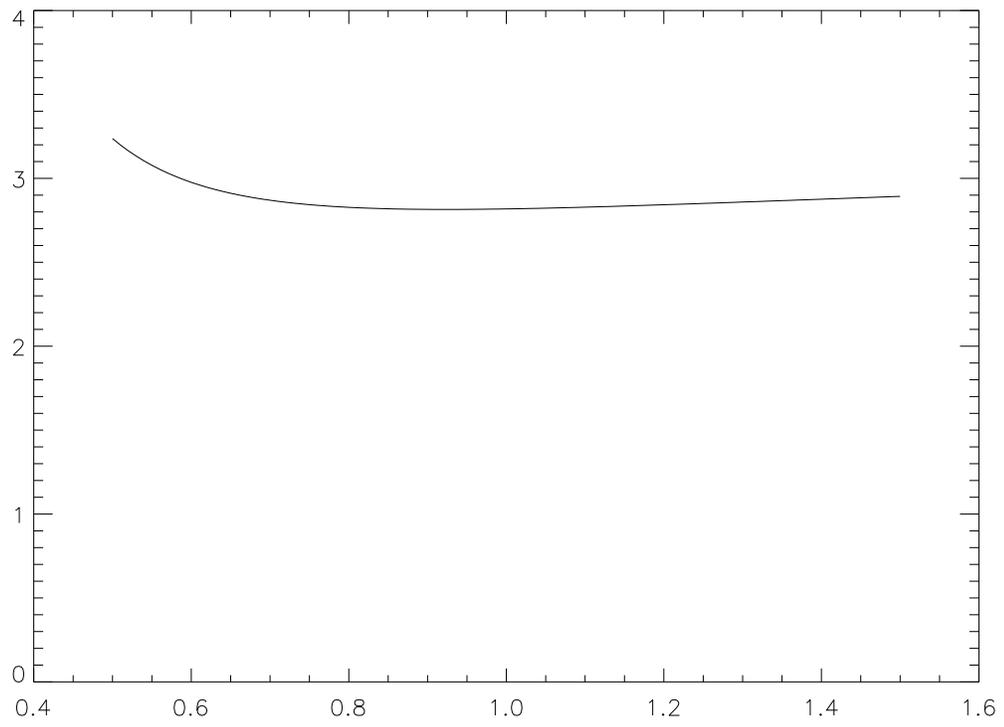}
\end{center}
\caption{This figure represents the dependence of  $H_0\eta_q$ on the parameter $w$. }
\label{fig5}
\end{figure*}

\begin{figure*}
\begin{center}
\includegraphics[width=15.0cm]{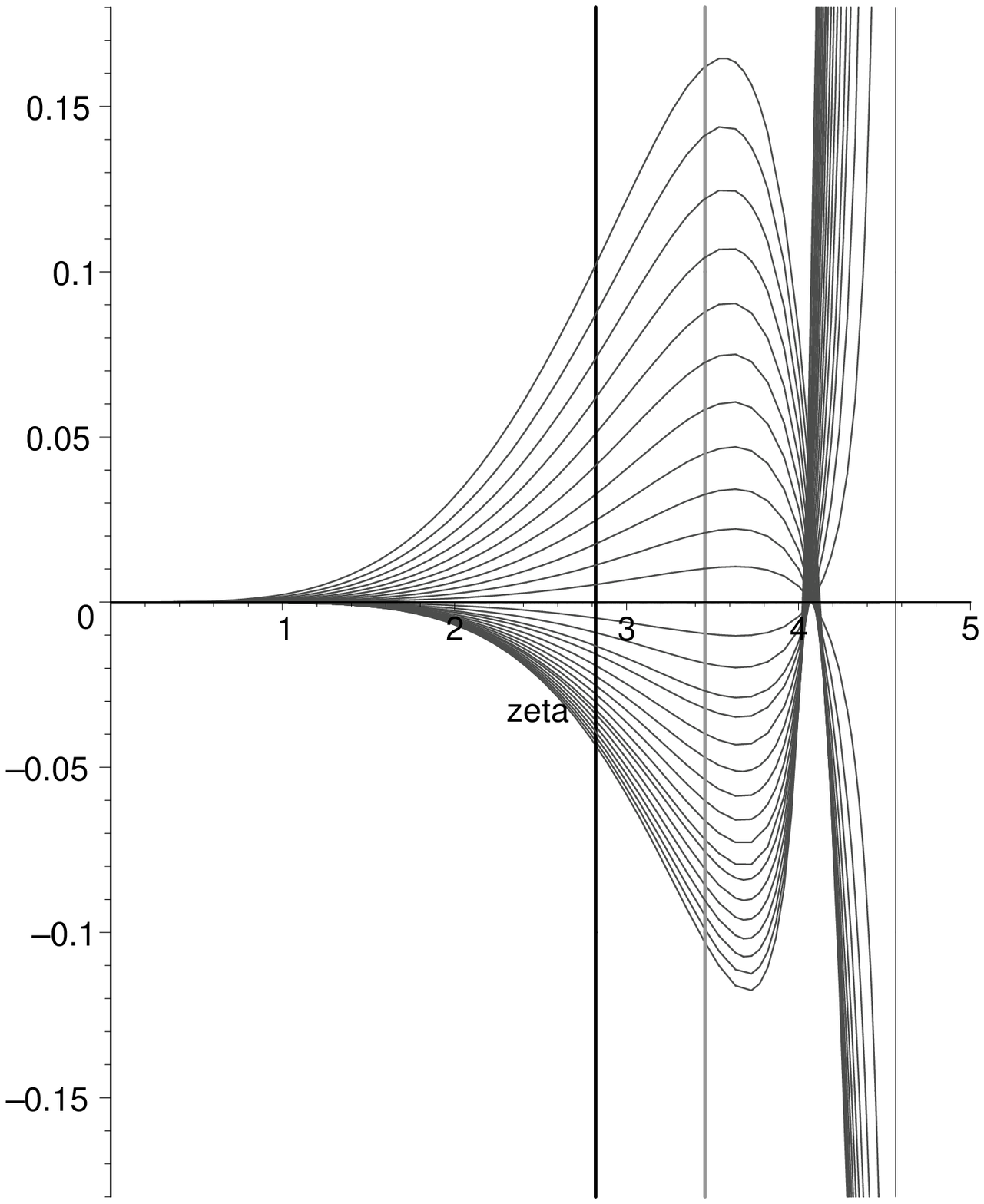}
\end{center}
\caption{This figure represents the function $\Delta$ of the variable $zeta = H_0\eta$. The thick vertical line represents the moment of time corresponding to the beginning of accelerated expansion of the Universe, while the next vertical line represents the current moment of time. It is seen that at the point of change from deceleration to acceleration, the value of $\Delta$ already exceeds 10\%. This means that for several problems, a Universe dominated by dark matter cannot be approximated by the standard model of the Universe. Instead, suitable approximations must be considered for each case.}
\label{fig6}
\end{figure*}


\begin{thebibliography}{25}

\bibitem{rei98}
Riess, Adam G.; Filippenko, Alexei V.; Challis, Peter; et al., 
Observational Evidence from Supernovae for an Accelerating Universe and a Cosmological Constant
The Astronomical Journal, Volume 116, Issue 3, pp. 1009-1038;
Riess, Adam G.; Filippenko, Alexei V.; Liu, Michael C.; Challis, Peter;
Tests of the Accelerating Universe with Near-Infrared Observations of a High-Redshift Type IA Supernova
The Astrophysical Journal, Volume 536, Issue 1, pp. 62-67.

\bibitem{sah00}
Sahni, V., Starobinsky, A.,
The Case for a Positive Cosmological $\Lambda$-Term
International Journal of Modern Physics D, Volume 9, Issue 04, pp. 373-443 (2000).


\bibitem{Padmanabhan:2002ji}
  T.~Padmanabhan,
  Phys.\ Rept.\  {\bf 380} (2003) 235
  [arXiv:hep-th/0212290].


\bibitem{Sahni:2004ai}
  V.~Sahni,
  Lect.\ Notes Phys.\  {\bf 653} (2004) 141
  [arXiv:astro-ph/0403324].


\bibitem{Copeland:2006wr}
  E.~J.~Copeland, M.~Sami and S.~Tsujikawa,
  Int.\ J.\ Mod.\ Phys.\  D {\bf 15} (2006) 1753
  [arXiv:hep-th/0603057].


\bibitem{Sahni:2006pa}
  V.~Sahni and A.~Starobinsky,
  Int.\ J.\ Mod.\ Phys.\  D {\bf 15} (2006) 2105
  [arXiv:astro-ph/0610026].

\bibitem{Frieman:2008sn}
  J.~Frieman, M.~Turner and D.~Huterer,
  arXiv:0803.0982 [astro-ph].

\bibitem{kam05}
V. Gorini, U. Moschella, A. Kamenshchik, and V. Pasquier
The Chaplygin gas, a model for dark energy in cosmology
AIP Conf. Proc. -- March 16, 2005 -- Volume 751, pp. 108-125
GENERAL RELATIVITY AND GRAVITATIONAL PHYSICS: 16th SIGRAV Conference on General Relativity and Gravitational Physics; 


\bibitem{Komatsu:2008hk}
  E.~Komatsu {\it et al.}  [WMAP Collaboration],
  arXiv:0803.0547 [astro-ph].


\bibitem{Sahni:2008xx}
  V.~Sahni, A.~Shafieloo and A.~A.~Starobinsky,
  arXiv:0807.3548 [astro-ph].

\bibitem{Xia:2008ex}
  J.~Q.~Xia, H.~Li, G.~B.~Zhao and X.~Zhang,
  arXiv:0807.3878 [astro-ph].


\bibitem{Caldwell:2002}
R.~R.~Caldwell,
Phys.\ Lett.\ B {\bf 545}, 23 (2002)

\bibitem{lib09}
Libanov, M. V.; Rubakov, V. A.; Sazhina, O. S.; Sazhin, M. V.
Phys.\ Rev.\ D, v. 79, id. 083521, 2009.


\bibitem{wmap}
http://lambda.gsfc.nasa.gov/product/map/dr4/best$_{-}$params.cfm

\bibitem{rub08}
Gorbunov, D.S., Rubakov, V. A. Vvedenie v teoriyu rannei Vselennoi. URSS, M., 2008.

\bibitem{lar11}
Larson D., Dunkley J., Hinshaw G., et al.
ApJ Supp., v.192:16, 2011.


\bibitem{con11}
Conley A., Guy J., Sullivan M., et al., 
astro-ph1104.1443, 2011.


\bibitem{san10}
Santos B., Garvalho J.C., Alcaziz J.S.
astro-ph1009.2733, 2010.


\bibitem{lin81}
Linde, A. D., Fizika elementarnikh chastitz i inflyatsionnaya kosmologia, M. Nauka, 1981. 

\bibitem{dol88}
Dolgov, A. D., Zel'dovich, Y. B., Sazhin M. V., Kosmologia rannei Vselennoi, M.: Izdatelstvo MGU, 1988. 

\bibitem{lib09}
Libanov, M. V.; Rubakov, V. A.; Sazhina, O. S.; Sazhin, M. V.,
CMB anisotropy induced by tachyonic perturbations of dark energy,
Physical Review D, vol. 79, Issue 8, id. 083521

\bibitem{kol94}
E. Kolb, M. Turner
The Early Universe. 1994
Westview Press. 

\bibitem{gr63}
Gradstein, I. S., Rizhik, I. M. Tablitsi integralov summ, ryadov i proizvedenii. Fizmatgiz, M. 1963. (Formula No. 3.166-22).

\bibitem{htf}
Batman, G. Erdelyi, A., HIGHER TRANSCENDENTAL FUNCTIONS, vol.1-3, MC GRAW-HILL BOOK COMP., INC. 1953.



\end{thebibliography}
\end{document}